\documentclass[11pt]{article}
\usepackage{amsmath,amssymb}
\usepackage{latexsym}
\def\beq{\begin{equation}}
\def\eeq{\end{equation}}
\def\bmul{\begin{multline}}
\def\emul{\end{multline}}
\def\dd{{\bar d}}
\def\cf{{\cal F}}
\def\tcf{\tilde{\cal F}}
\def\cg{{\cal G}}
\def\tcg{\tilde{\cal G}}
\begin{document}
\title{Generalized Geometry and M theory.}
\author{David S. Berman\\Department of Physics,\\Queen Mary\\Mile End Road\\
London E4 9NS,\\England\\ \\
Malcolm J. Perry\\
Department of Applied Mathematics and Theoretical Physics,\\
Centre for Mathematical Sciences,\\
University of Cambridge,\\
Wilberforce Road,\\
Cambridge CB3 0WA,\\ England\\ \\ and \\ \\ 
Perimeter Institute for Theoretical Physics,\\31 Caroline St N.,
\\Waterloo,\\Ontario N2L 2Y5,\\Canada.}
\maketitle
\begin{abstract}
We reformulate the Hamiltonian form of bosonic eleven dimensional supergravity
in terms of an object that unifies the three-form and the metric. For the 
case of four spatial dimensions, the  duality group is manifest 
and the metric and 
C-field are on an equal
footing even though no dimensional reduction is required for our results
to hold. 
One may also describe our results using the generalized geometry that emerges 
from membrane duality. The relationship between the twisted 
Courant algebra and the gauge symmetries of eleven dimensional supergravity 
are described in detail.

\end{abstract}

\newpage

\section{Introduction}

It has been a long standing puzzle that the spacetime metric and 
bosonic matter fields in M theory (or string theory) appear to be on different 
footings. The metric provides us with a notion of spacetime.
The matter fields are things that propagate in spacetime. In a theory
that ought to unify all fields, it seems strange that they are 
thought of as so distinct. It is the case that the Kaluza-Klein picture goes 
some way towards resolving this issue, however in M-theory one is still 
left with a
three form potential describing some bosonic matter whilst the metric 
describes the eleven dimensional spacetime.

In 1980, Julia discovered that 11-dimensional supergravity 
when compactified on tori of various 
dimensions, exhibited a host of symmetries \cite{Julia:1980gr}, \cite{Julia:1982gx},
which were enumerated by Morel and Thierry-Mieg \cite{TM:1980}, and by Eugene Cremmer
in the special case of dimension five \cite{Cremmer:1980gs}. They are also related to the 
symmetries of the vacuum Einstein equations discovered by Gibbons and Hawking, \cite{Gibbons:1979xm}.
They are now usually rather plainly referred to as M theory dualities and it
is well known that they are controlled by some Lie group in low dimensions.
In dimensions greater than $8$ it seems that some more complicated 
object is involved.
In 1986, de Wit and Nicolai found some tantalizing clues that these
duality symmetries had content beyond what one expects from a theory
compactified on a torus. They speculated that the duality groups somehow 
controlled the theory in a fundamental way, \cite{NdeW}.
The challenge of finding the true meaning of these groups in M-theory 
has subsequently fascinated  many authors resulting in copious 
publications \cite{msymm}. A particular mention should be made of \cite{hillmann} which has similar goals to those described in section 6 of this paper. \footnote{We apologize to those we have not cited,
it is our ignorance. If you feel that some work should be cited, please
let us know and we will gladly cite relevant works in the 
published version of this paper.} 

The present work is devoted to a reformulation of eleven dimensional 
supergravity (or at least its bosonic sector) to exhibit a duality symmetry 
without the use of any compactification. The duality symmetry
is an organizing principle and places both the metric and three form 
in representations of the duality group itself. The duality group also
controls the dynamics of the theory in a fundamental way. In some ways, the 
final version of the theory is rather like a $\sigma$-model for some
symmetric space related to the duality group. However there are fundamental
differences which are described towards the end of this paper.
On the route to our final result we make use of the generalized geometry 
recently described by Hitchin \cite{hitchin} and by 
Gualtieri \cite{Gualtieri:2003dx} and discussed 
in an M-theory context by Hull \cite{Hull:2007zu} and by Pacheco and 
Waldram \cite{Pacheco:2008ps}. We describe  the dynamics 
of the theory using this  generalized geometry. We also discuss
how the generalized geometry arises from the world volume description of the
M theory two-brane and from the symmetry algebra of M-theory itself.

We work with the specific case of the $SL(5)$ duality 
group. This is the group that would arise in a $T^4$ spatial reduction 
of M-theory. 
One begins in eleven dimensions with a local $SO(10,1)$ group and a 
global $GL(11)$ associated with the spacetime metric. 
We then wish to consider the subgroup of these spacetime groups
which is expected to 
be a local $SO(4)$ and a global $SL(4)$ corresponding to the four dimensions
of space. What happens however seems like a miracle, as there 
is a symmetry enhancement and the $SL(4)$ is enlarged to an $SL(5)$,
the local $SO(4)$ becomes $SO(5)$ and these groups act on both the metric
and the three-form in a unified way. 
It is this $SL(5)$ subgroup that we will study in detail. 
Let us emphasize that in this construction, no dimensional reduction 
has been imposed and all the fields can depend on the coordinates associated 
with these four spatial directions.

We suspect that one can extend our treatment to all of the
exceptional duality groups that have been found via dimensional
reduction. The first higher dimensional groups one encounters however 
will require the use
of fivebranes and a more complicated generalized metric structure
involving $C_6$, the six form potential corresponding to dual seven form
field strength of M theory. 
If one progresses to dimension seven, then even more 
complicated objects must be involved. We expect to address this in future work.

We use the Hamiltonian form of the theory, following closely the
work of  Dirac \cite{dirac}; Arnowitt, Deser and Misner \cite{canons}; and deWitt \cite{DeWitt:1967yk}. The reason we choose to work with the 
Hamiltonian form is because we wish to keep time separate.
Were it the case that one of the directions that are acted on by the
duality transformations were timelike, so that
signature of the metric was not Euclidean, then the duality would be of a 
more exotic type and the action of the symmetry on the C-field would imply a
complexification of the fluxes \cite{Hull:1998ym}. To avoid such difficulties,
we demand that time is treated separately this is what 
the canonical Hamiltonian form of gravity described by Dirac \cite{dirac}; Arnowitt, Deser and Misner \cite{canons}; and deWitt \cite{DeWitt:1967yk} accomplishes.

Our aim is to construct
the  Hamiltonian form of M theory using the metric and three
form data restructured in a way that makes the $SL(5)$ group manifest
and in some sense geometric. By geometric we mean the notion of 
generalized geometry
introduced by Hitchin \cite{hitchin} and others \cite{Gualtieri:2003dx}. 
In generalized stringy
geometry, the tangent space is complemented by a copy of the cotangent space at
each point in space. By so doing,  the $O(d,d)$ T-duality group becomes
natural, 
\cite{doubledform}. The appearance of
the cotangent space is associated with the geometry of the space as
seen by the string winding modes. The Kalb-Ramond field is the {\it{twist}}
that mixes the two halves of big space. One
half of the space is the usual geometry and the other is its  T-dual.
Of course one also must define a
{\it section} on the doubled space to relate it to a particular duality
frame in order to retrieve some kind of conventional spacetime picture.
For the $O(d,d)$ group of T-duality, this process is like picking
a polarization on the space. The string world sheet view of this is described
in \cite{doubledform} and the quantum consistency is discussed 
in \cite{berman1,berman2}.

The generalized geometry for M-theory is a bit different. We supplement
the tangent space with the space of two-forms
corresponding the space  seen by wrapped membranes. The three-form field
$C_{abc}$  is then the {\it{twist}} that mixes
the spaces. The space is no longer simply doubled since the number of
winding modes is different to the number of space dimensions for membranes. 
For the specific case at
hand, where the number of space dimensions is 4, the number of possible
windings is 6. Thus the enlarged space that includes the usual space
and its dual is 10 dimensional. Some work relating generalized geometry and branes has appeared in \cite{bonelli}.

We show how this generalized geometric structure arises directly from
considering the membrane world volume and its dual formulation pioneered by
Duff and Lu \cite{Duff:1990hn}.

Finally, armed with the geometric package from the membrane, 
we will construct a dynamical theory out of
the generalized metric that is equivalent to the canonical
Hamiltonian form of eleven dimensional supergravity. In doing so we 
will not assume that the theory is independent of any spacetime
directions. That is, we do  not carry out a dimensional reduction and we do
not assume the existence of any Killing vectors.

The only assumption that we do make to show equivalence to the usual
theory is to assume that the fields are independent of the 
dual {\it{wrapped}} membrane
directions. This assumption is a kind of {\it section} condition describing
how one takes a section through the extended space that results in 
our usual spacetime view.

We expect that this is not the only possible section  condition. In 
the doubled formalism of the string \cite{doubledform} there are many possible
ways of taking a section through the doubled space and these are of
course related by T-duality transformations. The absence of a globally
defined section is also possible and it is this possibility that gives
rise to so called nongeometric backgrounds such as T-folds where local
geometric patches are related by transition function given by
T-duality transformations. 
We will not deal these with issues here but note that the
possibility of a non-globally defined section in our enlarged space 
should lead to a
geometric construction of {\it{nongeometric}} U-folds. Recent work using 
generalized geometry and U-folds has appeared in \cite{Aldazabal:2010ef}.
We expect to return to this question in a future paper.

We will also ignore global considerations throughout the
analysis. Such global questions will be important in topologically
nontrivial situations and again we leave this for further work. A related
issue is that the groups described here will be real valued. Whilst this
makes sense as a solution generating symmetry, we expect that  only some
discrete arithmetic subgroup of the duality group is expected to be an M theory
symmetry, \cite{Witten:1995ex},\ \cite{Hull:1994ys}.

The structure of the paper is as follows. After carrying out the
canonical treatment of the bosonic sector of eleven dimensional
supergravity we determine the generalized metric that arises in
M-theory from examining the world volume dualisation of the
membrane. Finally, the dynamics of this generalized metric are
described and matched to the canonical form of the usual supergravity
under the condition that the metric is
independent of dual membrane winding mode directions. Along the way we
 will observe the
relationship between the Courant bracket algebra, the gauge symmetries 
in eleven dimensional supergravity and the generalized geometry
introduced by considering membrane duality transformations.
 
This paper covers a large area of research and as such there are many 
relevant and interesting works worthy of mention 
amongst these are \cite{general}.

\section{The Canonical Treatment of General Relativity}

The starting point is a canonical treatment of the bosonic part of M-theory.
The gravitational part of the action is just the usual Einstein-Hilbert term, 
an integral of the Lagrangian density $L$ over the spacetime ${\cal M}$. We follow the well worn path explored by Dirac \cite{dirac}, Arnowitt, Deser, Misner \cite{canons} and deWitt \cite{DeWitt:1967yk}.
In $d$ spacetime dimensions with metric tensor $g_{ab}$, we take this to be 
\beq
I_{grav}= \int_{\cal M} d^dx\ \, L \, , \quad{\rm{with}} \quad  L = \sqrt{g} R
\eeq
where $g=-\det{g_{ab}}$ and $R$ is the Ricci scalar (of the torsion-free
metric connection) of $g_{ab}$. Next, we make the mild topological restriction
that the spacetime can be foliated by a family of surfaces of constant time,
$\Sigma(t),$ where $t$ is a time co-ordinate. Let  $x^i$ be the 
$\dd=d-1$ spatial co-ordinates. Now we can
decompose the spacetime metric $g_{ab}$ into a purely spatial metric of
positive definite signature $\gamma_{ij}$, a lapse function
$\alpha$ and a shift vector $\beta^i$ by writing
\beq
g_{ab} = \begin{pmatrix}
 -\alpha^2+\beta_i\beta^i &\beta_j\\ \beta_i & \gamma_{ij} \end{pmatrix}.
\eeq
The inverse spacetime metric then becomes
\beq
g^{ab}= \begin{pmatrix} {-1/\alpha^2} & {\beta^j/\alpha^2}\\
{\beta^i/\alpha^2}&\gamma^{ij}-{\beta^i\beta^j/\alpha^2}
\end{pmatrix}.
\eeq
Using the above decomposition of the metric, the volume measure on spacetime
$d^dx\,\sqrt{g}$ is then given by $d^\dd x\,dt\,\alpha\sqrt{\gamma}$ 
where now $\gamma=\det\gamma_{ij}$.
The second fundamental form on the surfaces $\Sigma(t)$ is 
\beq
K_{ij}=\frac{1}{2\alpha}\bigl(D_i\beta_j + D_j\beta_i - \dot\gamma_{ij}\bigr).
\eeq
where now $D$ is the covariant derivative operator formed from 
$\gamma_{ij}$, and dot denotes the 
partial derivative with respect to $t$.

Given this decomposition into space and time, we can rewrite the gravitational
action as
\beq
I_{grav}= \int_{\cal M} d^{\dd} x\, dt\ \alpha\gamma^{\frac{1}{2}}
\bigl(R(\gamma)
+K_{ij}K^{ij} - K^2\bigr)
\eeq
$R(\gamma)$ is the Ricci scalar  
formed from the spatial metric $\gamma$, and $K$, the trace of the second 
fundamental form, is $K=K_{ij}\gamma^{ij}$. In the derivation of the above 
expression 
for $I_{grav}$,
boundary terms have  been discarded. \footnote{Throughout this work
we have neglected surface terms. These terms are potentially 
important in
topologically nontrivial situations and we leave such analysis for the
future.}

We now proceed to find the Hamiltonian for this theory. 
Firstly we must identify the canonical momenta conjugate to the
fundamental variables of the theory. These we denote by $\pi_\alpha,\pi^i_\beta$
and $\pi^{ij}_\gamma$ which are conjugate to $\alpha, \beta_i$ and $\gamma_{ij}$
respectively. 
One immediately finds two primary constraints, $\pi_\alpha\approx 0$
and $\pi^i_\beta\approx 0$, the momenta conjugate to the
lapse and the shift vector both vanish weakly.
The momentum conjugate to the spatial metric is given by
\beq
\pi_\gamma^{ij} = \gamma^{1/2}\Bigl(-K^{ij}+\gamma^{ij}K\Bigr).
\eeq
It should be noted that all momenta are densities of weight one 
with respect to the spatial metric $\gamma_{ij}$.
The Hamiltonian is then
\beq
H = \int_{\Sigma(t)} d^\dd x\ \bigl( \pi_\alpha \dot\alpha + 
\pi_\beta^i \dot\beta_i
+\pi_\gamma^{ij} \dot\gamma_{ij} - L \bigr)
\eeq
where $L$ is the Lagrangian density.
Requiring the primary constraints to be preserved under time evolution
results in two secondary constraints, the diffeomorphism constraint
\beq
\chi^i =-2D_j\pi^{ij}
\eeq
and the Hamiltonian constraint
\beq
{\cal H} = \gamma^{-1/2}\bigl(\pi^{ij}\pi_{ij} - \frac{1}{\dd-1}\pi^2 - 
\gamma R\bigr),
\eeq
both of which are also required to vanish weakly.
The Hamiltonian can now be written as
\beq
H = \int d^\dd x \ \bigl( \pi_\alpha\dot\alpha + \pi_\beta^i\dot\beta_i
+\alpha{\cal H}+\beta_i\chi^i \bigr)  .
\eeq
It should be noted that the complete Hamiltonian weakly vanishes. 
Each of these 
constraints is first class, and so corresponds to a gauge degree of 
freedom. Each constraint must be supplemented by a gauge condition
so that the Poisson bracket of all of the constraints and gauge conditions
has non-vanishing determinant. A consequence of the gauge fixing is that
Hamiltonian evolution of the system will preserve not only the constraints
but the gauge conditions. Thus imposing them at one instant will ensure that
they are imposed for all time. 

It is worth our while counting up the physical degrees of freedom to check
that they come out correctly. The lapse, shift and spatial metric together 
with their conjugate momenta give $2 + 2\dd +  \dd(\dd+1)$ degrees
of freedom per point in space. The constraints give $1 + \dd + 1 +\dd$
for $\pi_\alpha$, $\pi_\beta^i$, ${\cal H}$ and $\chi^i$ respectively and the 
same for each gauge condition. This leads to the physical phase space having 
dimension $(\dd+1)(\dd-2)$ at each point in space, as one expects 
for the Einstein
gravity.

Let us now study what this Hamiltonian means for the gravitational field. 
Firstly we are free to make a gauge choice for the lapse and shift. 
We can choose the synchronous gauge $\alpha=1$ and $\beta^i=0$ for 
definiteness and simplicity, but it does not really matter what choice 
we make. There are still $\dd+1$ gauge conditions that need to be chosen to 
restrict $(\gamma_{ij},\pi^{ij})$. 
Now look at the Poisson bracket algebra of the diffeomorphism and
Hamiltonian constraints. 
\beq
\{{\cal H}(x),{\cal H}(x^\prime\}= 
(\chi^i(x)+\chi^i(x^\prime))D_i\delta(x,x^\prime),
\label{eq:PBHHG}
\eeq
\beq
\{\chi_i(x),{\cal H}(x^\prime)\} = {\cal H}(x)D_i\delta(x,x^\prime)
\eeq
and
\beq
\{\chi_i(x),\chi_j(x^\prime)\}=\chi_i(x^\prime)\partial_j\delta(x,x^\prime)
+\chi_j(x)\partial_i\delta(x,x^\prime)
 \eeq
The algebra shows that if one imposes the Hamiltonian constraint everywhere, 
the diffeomorphism constraint is automatically satisfied. 
That comes about because if the Hamiltonian constraint is satisfied, then 
the Poisson bracket of the Hamiltonian with itself must also vanish, and 
hence the diffeomorphism constraint is also satisfied by virtue of 
(\ref{eq:PBHHG}), as was first demonstrated by
Moncrief and Teitelboim, \cite{Moncrief:1972cx}.
For this reason, we will focus only on the Hamiltonian constraint.
the Hamiltonian constraint contains two pieces, 
a kinetic piece that is quadratic in the momenta and 
a potential term that is proportional to the curvature.

The kinetic piece can be written
\beq
T= \mathfrak{G}_{ijkl}\pi^{ij}\pi^{kl}
\eeq
with
\beq
\mathfrak{G}_{ijkl}=\frac{1}{2}\gamma^{-{1/2}}
\biggl(\gamma_{ik}\gamma_{jl}+\gamma_{il}\gamma_{jk}
-\frac{2}{\dd-1}\gamma_{ij}\gamma_{kl}\biggr),
\eeq
and a potential piece 
\beq
V = \gamma^{1/2}R.
\eeq
If $T$ were the only term in the Hamiltonian, then each point in space 
would behave independently. The constraint would then be what one expects for 
the Hamiltonian of free particle motion on a space $M$ of dimension 
$\frac{1}{2}\dd(\dd+1)$ at each point $x$ in space.
The momenta $\pi^{ij}$ behave as components of a  covariant vector in 
this space, each pair of indices $i\le j$ behaving as one index in $M$. 
$\mathfrak{G}_{ijkl}$ thus is the inverse metric on the space $M$.
The metric on $M$, $\bar{\mathfrak{G}}^{ijkl}$ is consequently determined by
\beq
\mathfrak{G}_{ijkl}\bar{\mathfrak{G}}^{klmn} = \frac{1}{2}(\delta_i^m\delta_j^n
+\delta_i^n\delta_j^m),
\eeq
to be 
\beq
\bar{\mathfrak{G}}^{klmn}= \frac{1}{2}\gamma^{1/2}\biggl(\gamma^{km}\gamma^{ln}
+\gamma^{kn}\gamma^{lm} - 2\gamma^{kl}\gamma^{mn}\biggr).
\eeq
It should be noted that the inverse metric on $M$ is not found by raising
the indices of $\mathfrak{G}$ using the spatial metric.
The signature of $\bar{\mathfrak{G}}^{ijkl}$ is 
$(+^{\frac{1}{2}(\dd+2)(\dd-1)},-)$ 
for $\dd>1$.
One can think of the metric on $M$ as  being a metric on the space of all
symmetric tensors at each point in space. Suppose one has 
a symmetric tensor $h_{ij}$ at some point in space. Then the norm of $h_{ij}$
is \beq
\| h \|^2 = \bar{\mathfrak{G}}^{ijkl}h_{ij}h_{kl}.
\eeq
If $h_{ij}$ were to be  proportional to the spatial metric 
tensor, $h_{ij}=\phi\gamma_{ij}$ 
then \newline  $\|h\|^2=-\gamma^{1/2}\phi^2 \dd (\dd-1)<0$ provided we 
are in spacetime dimension $d>2$. It is only for $d>3$ that there is any 
meaningful idea of dynamical gravitation. Alternatively one can think of
the metric components $\gamma_{ij}$ as providing a set of co-ordinates 
on $M$. There  is a single negative (timelike) direction that corresponds
to conformal transformations of the spatial metric $\gamma_{ij}$.
The fact that $M$ does not have a positive definite metric is 
related to bad behavior of the Euclidean path integral where it has been found
that the Euclidean Einstein action is unbounded below, \cite{Gibbons:1978ac}.

We need to explore the geometry of $M$, so firstly define on $M$ a co-ordinate 
\newline $\zeta=4\sqrt{(\dd-1)/\dd}\gamma^{1/4}$.
$\zeta$ is timelike. Suppose that there are a further set of co-ordinates
$\xi^A, (A=1, \ldots, \frac{1}{2}(\dd+2)(\dd-1))$ orthogonal $\zeta$. Then the 
metric on $M$ can be written as
\beq
G_{AB}=\begin{pmatrix} -1&0\\0&\frac{\dd}{16(\dd-1)}\zeta^2 \bar G_{AB} 
\end{pmatrix}
\eeq
where $\bar G_{AB}$ is a metric on $\bar M,$ the surfaces of constant
$\zeta$. Writing the metric in this form shows us that $M$ is a
lorentzian cone over $\bar M$. The scale of $\bar M$ is set by $\zeta$.
One can calculate 

the curvature of $\bar M$. One finds that $\bar M$ is an Einstein space 
of negative curvature, with
\beq
\bar R_{AB} = -\frac{\dd}{2}\bar G_{AB}.
\eeq
One also finds that the Riemann tensor is 
covariantly constant.  Therefore $\bar M$ is  locally a symmetric space
$SL(\dd)$ acts transitively on $\bar M$, 
and the isotropy group is $SO(\dd)$, Thus $\bar M$ can be identified with 
coset space $SL(\dd)/SO(\dd)$. One may also easily calculate the curvature
of $M$. The Ricci scalar of $M$ is a constant times $\zeta^{-2}$. 
Thus $M$ is singular at 
$\zeta=0$ where the scale of the embedded $\bar M$ vanishes.

Now we construct the evolution equations in the synchronous gauge. The
Hamiltonian is 
\beq
H = \int d^{d-1}x\ {\cal H} = \int d^{\dd}x \ \biggl( \mathfrak{G}_{ijkl}
\pi^{ij}\pi^{kl} - \sqrt{\gamma}R \biggr).
\eeq
The evolution equations are now found by computing the appropriate Poisson
brackets. Firstly, the evolution of the metric is 
given by
\beq
\dot\gamma_{ij}  = \{\gamma_{ij},H\} = 
\frac{1}{2} \gamma^{-{1/2}}\biggl(\pi_{ij} -\frac{1}{\dd-1}\gamma_{ij}\pi\biggr).
\eeq
Recalling the definition of the momentum in terms of the second fundamental
form, we easily see that this equation just amounts to a restatement of the 
definition of momentum. The evolution equation for the momentum is a little
more complicated.
\beq
\dot\pi^{ij}=\{\pi^{ij},H\}=
\frac{1}{2}\gamma^{ij}\mathfrak{G}_{klmn}\pi^{kl}\pi^{mn}
-2\gamma^{-{1/2}}( \pi^{ik}\pi^j{}_k-\frac{1}{\dd-1}\pi\pi^{ij}
-\gamma^{1/2}(R^{ij}-\frac{1}{2}\gamma^{ij}R )).
\eeq
The last term involving the Einstein tensor comes from the
potential energy term and is only the only effect the potential energy
term has on the evolution.
We can rewrite these two equations as an evolution equation for the metric.
\beq
\ddot\gamma_{ij}-\dot\gamma_{ik}\dot\gamma_{jl}\gamma^{kl}
+\frac{1}{4(\bar d-1)}\bigl(\gamma^{kn}\gamma^{lm}-\gamma^{kl}\gamma^{mn}\bigr)
\dot\gamma_{kl}\dot\gamma_{mn}\gamma_{ij}=-\frac{1}{2}\dot\gamma_{ij}\gamma^{mn}
\dot\gamma_{mn}-\frac{1}{2}\bigl(R_{ij}-\frac{1}{2(\bar d-1)}R\gamma_{ij}\bigr).
\label{eq:geo}
\eeq

(\ref{eq:geo})\ is rather like the geodesic equation. The last term on the right
hand side is a kind of force coming from the curvature. If it were absent,
(\ref{eq:geo}) \ would be the geodesic equation for $\gamma_{ij}$ in the 
space $M$ at each point in physical space. We note that $t$ is not an 
affine parameter which is the reason for the first term on the right 
hand side of the equation. One may easily parameterize the curve in $M$ so
that it is affinely parameterized. Let $s$ be an affine parameter along the 
curve. It is related to $t$ by
\beq
\frac{\ddot s}{\dot s} = -\frac{1}{2}\gamma^{ij}\dot\gamma_{ij}=
-\frac{1}{2}\dot{\det\gamma} 
\eeq
One can now explicitly solve this equation to discover that
\beq
\dot s = \frac{{\rm constant}}{\zeta^2}
\eeq
where $\zeta$ is the timelike coordinate on $M$.
Since $\zeta=\lambda\gamma^{1/4}$ we see that time is essentially
specified by the volume of space, a fact familiar to cosmologists.
Each point in space would have a metric that
evolves independently of every other point. This is the explanation of the 
billiards picture.  The curvature term however has the effect of coupling 
different points in space to each other.

It should be noted 
that if  one compactified space on a torus, then of course
there is no spatial curvature. Additionally, each point in space is then
equivalent to every other point and so the evolution of the spacetime is
given by a single equation. The evolution equation is then just that of a 
particle moving in the symmetric space $\bar M$. The motion of such a particle
is integrable. Different solutions of the geodesic equation are then 
related to each other by $SL(\bar d)$ transformations which demonstrates
the origin and nature 
of the duality group for general relativity. The duality group 
for general relativity was first 
discovered in a somewhat different context by Buchdahl \cite{Buchdahl:1956zz}, 
and by Ehlers \cite{Ehlers:1957zz}
and discussed
by Geroch \cite{Geroch}.

Of course, if curvature is introduced, it seems as if our simple picture 
will be spoilt by the fact that different points in space obey 
different equations as the curvature will depend where in space one 
is.

Nevertheless, the observation that $\bar M$ is the same as the general 
relativity duality group motivates us to look further and see if it is
possible to formulate similar ideas in $M$ theory.

One can also construct a Lagrangian formulation for the evolution equation.
It is found by evaluating the canonical momentum in terms of the metric and 
switching the sign of the potential term. Thus the Lagrangian is the relatively
simple looking expression
\beq
I = \int d^{\bar d}x\,dt 
(\bar{\mathfrak{G}}^{ijkl}\dot\gamma_{ij}\dot\gamma_{kl}
+\sqrt{\gamma}R)
\eeq

Suppose one makes a global $SL(\bar d)$ transformation on the metric, so
that 
\beq
\gamma_{ij} \rightarrow \gamma_{ij}^\prime  = M_i{}^k M_j{}^l\gamma_{kl}
\eeq
with $M \in SL(\bar d)$. The action is then invariant. Since the action is 
not scale invariant, it is not possible to extend this symmetry 
to $GL(\bar d)$. Part of our aim is generalize this observation to
M theory.

\section{The Canonical treatment of M-theory}

In M-theory, a three form abelian gauge field is coupled to gravity and the 
system is then treated in an entirely similar way.  The new part of the  
action  consists 
of two pieces; the standard kinetic term $I_{minimal}$
which is defined in any number of spacetime dimensions
and a Chern-Simons like term $I_{CS}$ which can only be defined in eleven 
spacetime dimensions.  
\beq
I_{minimal} = -\int_{\cal M} d^dx\,\sqrt{g}\ \frac{1}{48}F_{abcd}F^{abcd}
\eeq
where the field strength $F_{abcd}$ is made from the exterior derivative
of a potential $C_{abc}$ where
\beq
F_{abcd} = 4\partial_{[a} C_{bcd]}
\eeq
$I_{minimal}$ has an abelian gauge invariance under which
\beq
C_{abc} \rightarrow C_{abc} + 3\partial_{[a} \Lambda_{bc]}
\eeq

The Chern-Simons term is 
\beq
I_{CS} = \lambda \int_{\cal M} d^dx \ \eta^{abcdefghijk}C_{abc}F_{defg}F_{hijk}
\eeq
where $\eta$ is the eleven-dimensional alternating tensor density. 
For M-theory, $\lambda$ is determined
by supersymmetry to be $\lambda =  2^{-7}3^{-4}$. It is possible to choose the 
opposite sign for this term but that is equivalent to changing the 
sign of $C_{abc}$\,; we have made the arbitrary choice for $\lambda$ to be
positive. The Chern-Simons term is 
also gauge invariant, but only up to boundary terms that are irrelevant 
to the discussions presented here. In what follows, in $d\ne 11$ spacetime 
dimensions we will take $\lambda =0$, but if $d=11$, then $\lambda \ne 0$.

 Just as for the
gravitational field, we split $C_{abc}$ up into its purely spatial
components and the remainder. Thus,
\beq
C_{abc} \rightarrow C_{ijk}\ \ \ \ {\rm and}\ \ \ \ C_{0ij}=B_{ij}.
\eeq
Then $C_{ijk}$ has a field strength, analogous to a magnetic field, of
\beq
F_{ijkl}= 4 \partial_{[i} C_{jkl]},
\eeq
and $B_{ij}$ has a field strength, analogous to an electric field, of
\beq
G_{ijk} = 3\partial_{[i} B_{jk]}.
\eeq
We now split up the action for the three-form theory using the 
space and time decomposition of both the metric and three-form fields to find
\begin{multline}
I  = \int_{\cal M }d^{\dd}x\,dt\,\alpha \gamma^{1/2}\ -\frac{1}{48}
\Bigl(F_{ijkl}F^{ijkl} - \frac{4}{\alpha^2}
\beta^i\beta^mF_{ijkl}F_m{}^{jkl}  - \frac{4}{\alpha^2}\dot C_{ijk}\dot C_{lmn}
\gamma^{il}\gamma^{jm}\gamma^{kn}\\ +\frac{8}{\alpha^2}\dot C_{ijk}G_{ijk} 
-\frac{4}{\alpha^2}G_{ijk}G^{ijk} 
+ \frac{8}{\alpha^2}\beta^l\dot C_{ijk}F_l{}^{ijk}
- \frac{8}{\alpha^2}\beta^lG_{ijk}F_l{}^{ijk}\Bigr)\\
 +\lambda\eta^{ijklmnpqrs}\Bigl(3B_{ij}F_{klmn}F_{pqrs} -
8C_{ijk}\dot C_{lmn}F_{pqrs}
+8C_{ijk}G_{lmn}F_{pqrs} \Bigr)
\end{multline}

The next step is to find the canonical momenta $\pi_B^{ij}$ conjugate 
to $B_{ij}$ and $\pi_C^{ijk}$ conjugate to $C_{ijk}$. We find 
a  primary constraint that $\pi_B^{ij} \approx 0$ as the Lagrangian
is independent of $\dot B_{ij}$. The momentum conjugate to $C_{ijk}$ is
\beq
\pi_C^{ijk}=\frac{\gamma^{1/2}}{6\alpha}\Bigl(
\dot C_{lmn}\gamma^{il}\gamma^{jm}\gamma^{kn}
-G^{ijk} -\beta^lF_l{}^{ijk}\Bigr) + 8\lambda\eta^{ijklmnpqrd}C_{lmn}F_{pqrs}
\eeq
In terms of the canonical variables, the Hamiltonian for the three-form field
is
\begin{multline}
H = \int_{\Sigma}d^{\dd}x  \,  \pi_B^{ij}\dot B_{ij} + 
3\alpha\gamma^{-{1/2}}\pi_{ijk}\pi^{ijk}
+\pi^{ijk}(G_{ijk}+\beta^lF_{lijk})
+\frac{\alpha\gamma^{1/2}}{48}F_{ijkl}F^{ijkl}\\
+\frac{\lambda}{54}\alpha\gamma^{-{1/2}}\eta_{ijklmnpqrs}\eta^{ijk}{}_{tuvwxyz}
C^{lmn}F^{pqrs}C^{tuv}F^{wxyz}\\
+\lambda\eta^{ijklmnpqrs}(-3B_{ij}F_{klmn}F_{pqrs} + 
8C_{ijk}\beta^tF_{tlmn}F_{pqrs}+ 48 C_{ijk} \pi_{lmn} F_{pqrs}      ).
\end{multline}
There is a new secondary constraint \footnote{We have made our conventions
consistent throughout the paper. One consequence is that various numerical 
factors look rather strange, in particular those associated with the 
normalization of the gauge constraint and in section five where 
the C field is discussed}
 resulting from the gauge invariance 
of the three-form potential
\beq
\chi^{ij}=-\sqrt{2}(\partial_k\pi^{ijk} 
+\lambda\eta^{ijklmnpqrs}F_{klmn}F_{pqrs}).
\eeq
This new constraint we will call the gauge constraint.

The gauge constraint is reducible, meaning that not all of its components are
independent. It obeys the identity 
\beq
\partial_i \chi^{ij}=0 \, .
\eeq

The Hamiltonian for the three-form theory is then
\begin{multline}
H_C = \int_{\Sigma} d^{\dd}x\ \pi_B^{ij}\dot B_{ij} + 
\frac{\sqrt{2}}{3}B_{ij}\chi^{ij}
+ 3\alpha\gamma^{-{1/2}}\pi^{ijk}\pi_{ijk} + \frac{1}{48}\alpha\gamma^{1/2}
F^{ijkl}F_{ijkl}\\ + \pi^{ijk}\beta^lF_{ijkl} 
+8\lambda\eta^{ijklmnpqrs}\beta^tC_{ijk}F_{tlmn}F_{pqrs}
\end{multline}

The Hamiltonian for the three-form can now be combined with the 
gravitational Hamiltonian to produce 
an expression for the complete bosonic part of M-theory.
\beq
H = \int_{\Sigma} d^{\dd}x \ \pi_\alpha\dot\alpha + \pi_\beta^i\dot\beta_i
+\pi_B^{ij}\dot B_{ij} + \beta_i\bar\chi^i + \alpha\bar{\cal H}
+\frac{3}{\sqrt{2}}B_{ij}\chi^{ij}.
\eeq 
The inclusion of the three-form has modified the diffeomorphism constraint,
$\chi^i\rightarrow\bar\chi^i$.
The modified constraint is 
\beq
\bar\chi^i = -2D_j\pi^{ij} + F^{ijkl}\pi_{jkl} - 
8\lambda\eta^{jklmnpqrst}F^i{}_{jkl}
C_{mnp}F_{qrst}
\eeq
Similarly, the Hamiltonian constraint is modified 
${\cal H}\rightarrow\bar{\cal H}$
so that the modified constraint is
\beq
\bar{\cal H} = \gamma^{-1/2}\Bigl(\pi^{ij}\pi_{ij}-\frac{1}{\dd-1}\pi^2
+3\pi^{ijk}\pi_{ijk} + \gamma(-R + \frac{1}{48}F^{ijkl}F_{ijkl})\Bigr).
\eeq

All of the equations can be made to look nicer by making the following
canonical transformation
\beq
\hat \pi^{ijk} = \pi^{ijk} -8\lambda\eta^{ijklmnpqrs}C_{lmn}F_{pqrs}
\eeq
so that constraints are now
\beq
\bar \chi^i = -2D_j\pi^{ij} + F^{ijkl}\hat\pi_{jkl}
\eeq
\beq
\bar{\cal H} = \gamma^{-1/2}\Bigl(\pi^{ij}\pi_{ij} - \frac{1}{\dd-1}\pi^2
+3\hat\pi^{ijk}\hat\pi_{ijk}+\gamma(-R + \frac{1}{48}F^{ijkl}F_{ijkl})\Bigr).
\eeq
and
\beq
\chi^{ij} = -\sqrt{2}(\partial_k\hat\pi^{ijk} 
+3\lambda\eta^{ijklmnpqrs}F_{klmn}F_{pqrs}).
\eeq
The Hamiltonian then remains the same except that one uses the new form for 
each of the constraints. It can now be seen that the overall structure of the 
theory is universal and  independent of whether the 
Chern-Simons term is present  or not.  The Hamiltonian still consists of two 
parts, a kinetic piece $T$ given by
\beq
T = \gamma^{-1/2}\Bigl(\pi^{ij}\pi_{ij} - \frac{1}{\dd-1}\pi^2
+3\hat\pi^{ijk}\hat\pi_{ijk}\Bigr),
\eeq
and a potential piece $V$ given by
\beq
\gamma^{1/2}\Bigl(-R + \frac{1}{48}F^{ijkl}F_{ijkl}\Bigr).
\eeq

We should check that we have the correct number of physical degrees of freedom
in our theory. The counting for the gravitational sector is unchanged.
For $B_{ij}$ and $C_{ijk}$ the number of degrees of freedom are 
$\frac{1}{2}\dd(\dd-1)$ and $\frac{1}{6}\dd(\dd-1)(\dd-2)$ respectively. 
Doubling this total results in a phase space that has dimension
$\frac{1}{3}\dd(\dd^2-1)$. From the vanishing of $\pi_B^{ij}$ and the 
conjugate gauge conditions, one subtracts $\dd(\dd-1)$.
>From the vanishing of $\chi^{ij}$ one might at first sight think that
there are a further $\frac{1}{2}\dd(\dd-1)$ constraints. However, the fact that
$\chi_{ij}$ is identically divergence-free means that there are in fact 
fewer independent 
equations here. The divergence-free condition seems to  subtract off 
$\dd$ from the number of conditions  since there are $\dd$ vector fields. 
However our counting 
is still not quite right 
as one of these $\dd$ conditions is redundant coming as it does from
a vector that is itself divergence free. So the correct number to subtract off
is $\dd-1$. There are therefore $\frac{1}{2}(\dd-1)(\dd-2)$ 
independent equations
in $\chi_{ij}=0$. Each of the
independent constraints is still first class and so must be supplemented by 
a gauge condition. One therefore subtracts off a further $(\dd-1)(\dd-2)$
to obtain the number of physical degrees of freedom. Thus, the physical
phase space of the theory has dimension $\frac{1}{3}(\dd-1)(\dd-2)(\dd-3)$.
The rather complicated degree of freedom counting will lead in a quantum
version of this theory to ghosts for ghosts and 
ghosts for ghosts for ghosts \cite{tm:brs} in addition to 
just the usual ghosts.

Now we will examine in detail the Poisson bracket algebra for three forms 
coupled to gravity. For convenience, we will put $\lambda=0$ in what follows.
the Poisson brackets for three-forms coupled to gravity have
been discussed in the literature previously by
Baulieu and Henneaux, \cite{Baulieu:1987pz}.
An involved calculation reveals
\beq
\{ {\bar{\cal{H}}}(x),{\bar{\cal {H}}}(x^\prime\}= 
(\bar\chi^i(x)+\bar\chi^i(x^\prime))D_i\delta(x,x^\prime),
\label{eq:PBHH}
\eeq
\beq
\{ \bar\chi_i(x),{\bar{\cal{ H}}}(x^\prime)\} = 
{\bar{\cal{H}}}(x)D_i\delta(x,x^\prime)+{\frac{\sqrt{2}}{18}}
\pi_{ijk}\chi^{jk}\gamma^{-1/2}\delta(x,x^\prime),
\eeq
\beq
\{\bar\chi_i(x),\bar\chi_j(x^\prime)\}=\bar\chi_iD_j\delta(x,x^\prime)
+\bar\chi_j(x)D_i\delta(x,x^\prime)+
\frac{1}{\sqrt{2}}F_{ijkl}\bar\chi^{kl}\delta(x,x^\prime), \label{eq:diffs}
\eeq
and
\beq
\{ {\bar{\cal{H}}}(x),\chi^{ij}(x^\prime)\}=
\{\bar\chi_i(x),\chi^{jk}(x^\prime)\}
=\{\chi^{ij}(x),\chi^{kl}(x^\prime)\}=0.\eeq

The first thing to note about the above algebra is that the argument of Moncrief
and Teitelboim \cite{Moncrief:1972cx} is still valid. 
If ${\bar{\cal {H}}}$ vanishes everywhere, then
the Poisson bracket of ${\bar{\cal {H}}}$ with itself vanishes everywhere 
which implies that the 
diffeomorphism constraint holds. Similarly the second Poisson bracket relation
then implies that the gauge constraint holds. Next, notice that the 
diffeomorphism constraint together with the gauge constraint is a 
sub-algebra of the entire algebra. Consequently, one expects this sub-algebra
to control the behavior of the fields in the theory under purely 
spatial diffeomorphisms and gauge transformations.

Gauge transformations of the three-form are easy to compute. One finds the
gauge transformation of any field by computing its Poisson bracket 
with $\int dx \xi_{ij}(x) \frac{3}{\sqrt{2}}\chi^{ij}(x).$ Thus one finds
the gauge variation of $C_{ijk}$ to be 
\beq
\delta C_{ijk} = \{ C_{ijk}(x),
\frac{3}{\sqrt{2}}\int dx^\prime \xi_{lm}(x^\prime)\chi^{lm}
(x^\prime) \} = 3\partial_{[i}\xi_{jk]} \eeq
If $\xi$ is exact, $\xi_{jk}=2\partial_{[j}\lambda_{k]}$ for some $\lambda$, then
the gauge transformation vanishes which reflects the fact that $\chi^{ij}$ is
reducible.  If one performs a gauge transformation on any of the fields using
and exact gauge parameer, then the variation is identically zero. 
Consequently, if one tries to determine the algebra of
symmetries of a theory just from looking at the action of the constraints on
the fields of the theory, as we have done here, there is a danger of missing
such terms in the resultant symmetry algebra describing the algebra
of gauge transformations.

One naively expects to find that under spatial co-ordinate transformations
generated by  vector fields $X^i$, a field $F$ will transform by a quantity 
related to the Lie 
derivative of $F$, by $\delta F = {\bf L}_X F$. 
One can carry out this calculation explicitly for our diffeomorphism constraint.
A co-ordinate transformation generated by $X^i$ is found by taking the 
Poisson bracket of any field with
$\int dx X^i(x)\bar\chi_i(x)$. Thus the variation of the metric is
\beq \delta\gamma_{ij}(x)=\{\gamma_{ij}(x),
\int dx^\prime X^k(x^\prime)\bar\chi_k
(x^\prime)\}=D_iX_j+D_jX_i \eeq
as one expects. The variation of $C_{ijk}$ is calculated in the same way 
and one finds
\beq \delta C_{jkl} = X^iF_{ijkl}\label{eq:varc} \eeq
(\ref{eq:varc})\ is not what one would have expected as it 
is a combination of the Lie derivative and a gauge transformation
\beq \delta C_{jkl} = ({\bf L}_X C)_{jkl} - 3\partial_{[j}\biggl(X^mC_{kl]m}
\biggr)\eeq
The final point we wish to emphasize about the constraint algebra can be 
seen by looking at the Poisson bracket of two diffeomorphism constraints.
It involves a term proportional to the four-form field strength times
the gauge constraint. such  a  term is somewhat unexpected as it  implies that
the algebra of symmetry transformations is field dependent. There is no 
problem in principle with such a situation but this kind of thing is likely 
to lead to difficulties with the Jacobi identity.

The end-point of this section is then to note that the structure of
the Hamiltonian constraint is rather similar to that for pure gravity
except that it contains both the three-form and the metric. However, it 
does not have any obvious structure involving any coset space, even
after the timelike direction in the kinetic term has been factored out.
In section six, we will return to this point and show that there is indeed
a hidden coset space construction, exactly paralleling the case for pure 
gravity.

\section{Duality for the M2-brane}

In this section we follow closely the work of Duff and Lu \cite{Duff:1990hn}. One of the fundamental constituents of M-theory is the $M2$-brane. Suppose 
that the $M2$ is embedded in a spacetime with metric $g_{ab}$ and 
three-form $C_{abc}$. The location of the $M2$ in spacetime is given by
$X^a(\xi)$ where $\xi^\mu$ are world-volume co-ordinates and the world-volume
metric is $h_{\mu\nu}$ of signature $(-++)$.
The bosonic part of the $M2$ action is
\beq
I= \int d^3\xi\ \sqrt{-h} \ \ 
\bigl(\frac{1}{2}h^{\mu\nu}\partial_\mu X^a
\partial_\nu X^b g_{ab} \ \ +
\ \ \frac{1}{6}\epsilon^{\mu\nu\rho}\partial_\mu X^a
\partial_\nu X^b \partial_\rho X^c C_{abc}- \frac{1}{2}\bigr). 
\label{eq:M2action}
\eeq
where $h=\det{h_{\mu\nu}}$ and $\epsilon^{\mu \\nu \rho}$ denotes the alternating tensor.
$\kappa-$symmetry for the supersymmetric extension of this action 
requires the $M2$  to propagate in a background spacetime that 
satisfies the classical equations of motion precisely as described in 
section two, 
\cite{Bergshoeff:1987qx}. 

The $M2$ brane exhibits a duality symmetry that is somewhat similar to
that found for the string. Suppose that the string is compactified on 
a $d$-dimensional torus, then it exhibits an $SO(d,d)$ symmetry.
This $SO(d,d)$ symmetry has been related to doubled geometry which
has given some insight into the nature of string theory.
In this section, we will examine duality symmetry for the $M2$-brane.
Suppose for the moment that  we compactify on a $d$-dimensional torus,
so that there are 
$d$ commuting Killing vectors. The metric and three-form
fields will be independent of the $X^a$ associated with these Killing vectors.
we require the above condition is in order to find the 
conventional duality symmetry.
Suppose in addition that there are no other directions in spacetime in which 
the $M2$-brane is moving. Under such simplifying assumptions, 
the equations of motion that follow from
the action (\ref{eq:M2action}) are 
\beq
h_{\mu\nu} = \partial_\mu X^a \partial_\nu X^b g_{ab} \label{eq:metric}
\eeq
for the world-volume metric and 
\beq
\partial_\mu \cg^\mu_a =0
\eeq
where
\beq
\cg^\mu_a = \bigl(\sqrt{-h}g_{ab}\cf_\mu^b + \frac{1}{\sqrt{2}}
C_{abc}\tcg^{\mu\ bc}\bigr),
\eeq
\beq
\cf_\mu^a = \partial_\mu X^a, \label{eq:FS}\eeq
and
\beq
\tcg^{\mu\ ab}=\frac{1}{\sqrt{2}}\sqrt{-h}\epsilon^{\mu\nu\rho}
\cf_\nu^a\cf_\rho^b. \label{eq:fsg}
\eeq
The last equation, (\ref{eq:fsg}), has vanishing divergence as a consequence
of (\ref{eq:FS}). Thus
\beq
\partial_\mu \tcg^{\mu\ ab} \equiv 0. \label{eq:BI}
\eeq
This equation is therefore the  Bianchi identity.  
The physical content of the Bianchi identity can be seen by finding its 
solutions. At least locally, they are given by
\beq
\cf_\mu^a = \partial_\mu X^a
\eeq
for some $X^a$. In fact, this last result will hold independently of whether
there are any Killing vectors or not. 

One can now construct a first order action for the $M2$-brane. 
The Lagrangian density  is 
\beq
L = -\sqrt{-h}\biggl(\frac{1}{2}h^{\mu\nu}\cf^a_\mu \cf^b_\nu g_{ab}
+\frac{1}{3}\epsilon^{\mu\nu\rho}\cf^a_\mu \cf^b_\nu \cf^c_\rho C_{abc}
-\partial_\mu X^a \bigl(h^{\mu\nu}\cf_\nu^b g_{ab}
+\frac{1}{2}\epsilon^{\mu\nu\rho}\cf^b_\nu\cf^c_\rho C_{abc}\bigr) +\frac{1}{2}
\biggr).
\eeq
The first order action is entirely equivalent to (\ref{eq:M2action}) as can
be seen by finding the $\cf^a_\mu$ equation of motion and solving it purely 
algebraically to find $\cf^a_\mu = \partial_\mu X^a$. Substituting this back 
into the first order action reproduces (\ref{eq:M2action}) exactly.

In the dual description we will exchange the roles of
Bianchi identities and equations of motion.
The following first order Lagrangian, is duality symmetric,
\beq
L = \sqrt{-h}\biggl(\frac{1}{2}h^{\mu\nu}\cf_\mu^a\cf_\nu^bg_{ab}
+\frac{1}{6}\epsilon^{\mu\nu\rho}\cf^a_\mu\cf^b_\nu\cf^c_\rho C_{abc}
+\frac{1}{2\sqrt{2}}\epsilon^{\mu\nu\rho}\partial_\mu y_{ab} \cf^a_\mu\cf^b_\nu
-\frac{1}{2}\biggr). \label{eq:DS}
\eeq
In the above expression $y_{ab},$ antisymmetric under $a\leftrightarrow b$, 
is a dual winding co-ordinate. 
The $y_{ab}$ equation of motion is now the Bianchi identity 
(\ref{eq:BI}). Solving it as we did earlier and substituting for
$\cf^a_\mu$ leads to the action (\ref{eq:M2action}) provided the background 
admits the appropriate Killing vectors so that $g_{ab}$ and $C_{abc}$ are 
independent of the co-ordinates. Thus provided the Killing vectors exist,
the duality symmetric action is equivalent to the original one.  
The equation of motion  for $\cf^a_\mu$ is now
\beq
h^{\mu\nu}\cf^b_\nu g_{ab} + \frac{1}{2}\epsilon^{\mu\nu\rho}\cf^b_\nu\cf^c_\rho
C_{abc} + \frac{1}{\sqrt{2}}
\epsilon^{\mu\nu\rho}\bigl(\partial_\rho y_{ab}\bigr)
\cf^b_\nu
=0.
\eeq

We can re-organize the information about the equations of motion and Bianchi
identities by defining 
\beq
\tcf_{\mu\ ab} = \partial_\mu y_{ab}
\eeq
and writing
\beq
\begin{pmatrix}
\cg_{\mu a}\\ \tcg_\mu^{mn}
\end{pmatrix} = \begin{pmatrix} g_{ab}+\frac{1}{2}C_a{}^{ef}C_{bef}&
\frac{1}{\sqrt{2}}C_a{}^{kl}\\ \frac{1}{\sqrt{2}}C^{mn}{}_b & g^{mn,kl}
\end{pmatrix}\begin{pmatrix}\cf_\mu^b\\ \tcf_{\mu\ kl}\end{pmatrix},  
\label{genmet}
\eeq
where $ g^{mn,kl}=\frac{1}{2}(g^{mk}g^{nl}-g^{ml}g^{nk})$ and has the
effect of raising an antisymmetric pair of indices. $\cf_\mu^a $ and 
$\tcf_{\mu\ ab}$ are straight derivatives of the co-ordinates and are therefore
rather like displacements, whereas $\cg_{\mu\ a}$ and $\tcg_\mu^{ab}$ 
are rather like field strengths. A generalized displacement can be defined
by 
\beq
\cf_\mu^M = \begin{pmatrix}\cf_\mu^a\\ \tcf_{\mu\ ab}\end{pmatrix}\eeq
and
a generalized field strength by
\beq
\cg_{\mu\ M} = \begin{pmatrix}\cg_{\mu\ a}\\ \tcg_\mu^{ab}\end{pmatrix}\eeq
The equation of motion and Bianchi identity can then both be written as 
\beq
\partial_\mu \cg^\mu_M = 0. 
\eeq
The field strengths are then related to the displacements by
\beq
\cg_{\mu\ M} = M_{MN}\cf^N_\mu.
\label{genmet1}
\eeq
$M_{MN}$ is our generalized metric, explicitly defined by equation
(\ref{genmet}). This is the generalization of the spacetime metric
to include the potential of the $3$-form potential to be found in M-theory. 

Another, perhaps more intuitive way of thinking of this, is to use 
the original definitions of $\cf_\mu^M$ and $\cg^\mu_M$ and re-write the
Lagrangian as
\beq
L = \cg^\mu_M \cf_\mu^M.
\eeq
This last form of the action should remind us of the action for magnetostatics
in which the role of the magnetic displacement $B$ is played by $\cf$
and the role of the magnetic field $H$ by $\cg$. The generalized metric
is then seen to be rather similar to that of the magnetic permeability.
The word ``magnetic'' has been used here because the $M2$-brane is presumed 
to only be moving in spacelike directions and only responds to magnetic
components of the field strength. If that were not the case, one could
just as well think of electrostatics and make the analogy with $D$ and $E$.

The nature of membrane duality is now apparent. Take the collection of
membrane one forms where one has both spacetime and winding co-ordinates:
\beq
dZ^M = \begin{pmatrix} dx^a\\ dy_{ab} \end{pmatrix} \eeq and make a co-ordinate 
transformation so that
\beq
dZ \rightarrow \hat dZ  = TdZ \eeq
for some constant matrix $T$. Then 
\beq L = dZ^T M dZ \eeq  gets mapped into
\beq L = dZ^T T^T M T dZ = dZ^T \widehat{M} dZ \eeq
where \beq \widehat{M} = T^TMT. \eeq
So as long as $T$ is invertible $\widehat{M}$ is equivalent to $M$.
This is what is meant by duality. 

To progress further one needs to specify the number of relevant 
directions since each case is different. The remainder of this paper
is devoted to this topic and to understanding how and in what way
the duality group has meaning even when no Killing vectors are present.

\section{Generalized Geometry}

In conventional Riemannian geometry, one has a differentiable manifold 
${\cal M}$ that has a set of co-ordinates together with a distance
function in the form of a  metric tensor $g_{ab}$. 
At each point of the manifold, one has the tangent space $T$
which is spanned by vectors. The metric allows one
to calculate the norm of these vectors.

Vector fields $X$ are then
sections of the tangent bundle of the manifold, $T({\cal M})$. 
Suppose that 
one  performs an infinitesimal transformation of the co-ordinates
$x^a \rightarrow x^a+X^a$. Under such a  transformations, a tensor $S$
will transform to $S+\delta S$ where
$\delta S =  {\bf L}_X S$ and ${\bf L}$ is the Lie derivative.
The diffeomorphisms when acting on tensors then form an algebra since
\beq
{\bf L}_{X_1}{\bf L}_{X_2} - 
{\bf L}_{X_2}{\bf L}_{X_1}={\bf L}_{[X_1,X_2]}.
\eeq

In generalized geometry, these structures are enlarged. 
Courant \cite{Courant}, Hitchin \cite{hitchin}
and Gualtieri \cite{Gualtieri:2003dx} have shown
how to enlarge the space to include $p$-forms. In their work,  the tangent 
space at 
each point is enlarged to include a $p$-form; the generalized tangent space 
being $T \oplus \Lambda^p(T^*)$. The generalized metric of the previous section
provides the manifold with a generalized metric structure for the 
specific case of $p=2$. Essentially, the metric has been replaced by
an object that contains the original metric together with a $3$-form
potential $C$ whose components are $C_{abc}$ and whose field strength is then  
$F=dC$.
>From here on, since we are focused on two-branes in M theory, we will 
only consider the case of $p=2$.
Higher dimensional examples occur when there are fivebranes are present 
and we must 
have a generalized structure of the 
form $T \oplus \Lambda^2(T^*) \oplus \Lambda^5(T^*)$. 
An analysis of this and other more complex situations will appear in a 
future paper. 
The generalized tangent bundle in the case at hand is
$T({\cal M})\otimes \Lambda^2 T^*({\cal M)}$. Sections of this  
bundle, $A$, are the sum of vector field $X$ and a two-form $\chi$,
$A = X \oplus \chi$. In generalized geometry, one must
supplement the co-ordinates $x^a$ by the winding
co-ordinates $y_{ab}$ introduced in the previous section. Diffeomorphisms
are found in the usual way. Reparameterizations of $x^a$ that are only  
$x$-dependent are the usual co-ordinate transformations. Reparameterizations
of $y_{ab}$ that are only $x$-dependent amount to gauge transformations of 
the $C$-field. Suppose one defines a generalized infinitesimal line element
$d\mathfrak{s}$ by
\beq
d\mathfrak{s}^2 = M_{MN}dZ^M dZ^N
\eeq
where $Z^M=(x^a,y_{ab})$.
Then if $y_{ab}
\rightarrow y_{ab}+\frac{1}{\sqrt{2}}\lambda_{ab}$ then demanding that
the generalized line element remains invariant,  induces a gauge transformation
$C_{abc} \rightarrow C_{abc} + \partial_{[a}\lambda_{bc]}$.
These are the usual gauge transformations to be expected of an abelian three 
form potential. However these gauge transformations are reducible. If 
$\lambda$ is exact, then since $d\lambda=0$, $C$ is unchanged. 
Any gauge transformation that is exact has  no effect on the fields of the 
problem. 
Composition of general transformations shows, following Courant, that
these transformations form a Lie algebroid rather than a Lie algebra.
This really just a way of saying that there is extra structure beyond the 
tangent bundle of the manifold; in our case, the space of two-forms.
Vectors together with two-forms constitute the gauge transformations 
of the theory we are interested in. The Courant bracket algebra for 
the composition of two generalized gauge transformations is
\beq
[X+\xi, Y+\eta ]_C  =  [X,Y] + {\bf L}_X\eta - {\bf L}_Y\xi 
-d(\iota_X\eta-\iota_Y\xi) \label{eq:diffq}
\eeq
where the bracket with a suffix $C$ is the Courant bracket, the bracket 
 with no suffix is the usual Lie bracket, and $X,Y \in T$ and $\eta,\xi \in
\Lambda^2(T^*)$. This is almost the same as the algebra  diffeomorphisms
found in the canonical version of M-theory.  The first term on the right in 
(\ref{eq:diffq}) is the usual Lie bracket on expects from commuting two 
diffeomorphisms. The second and third terms are what one expects from carrying
out a diffeomorphism on the two-forms. The last term is an extra 
contribution to the gauge transformations of the three form. However, 
since it is exact, it has no effect on the physical fields of the theory. 
 The only discrepancy between 
(\ref{eq:diffs})  and (\ref{eq:diffq}), is thus a  term involving the field 
strength.

Suppose now that we take a section of the generalized tangent bundle
and twist it by the three-form. Let $\rho_C$ be the twist operation. 
\beq
(X,\xi) \rightarrow \rho_C(X,\xi) = (X,\xi+\frac{1}{\sqrt{2}}\iota_X C)
\eeq
The idea of the twist operator is to take a background and add some $C$-field
to it. Thus \beq
\rho_{C_1}\rho_{C_2} = \rho_{C_1+C_2}\eeq
since $C$ is an abelian field.
Then
\beq
[\rho_C(X,\xi),\rho_C(Y,\eta)]_C = \rho_C([X+\xi,Y+\eta]_C + 
\frac{1}{\sqrt{2}}\iota_Y\iota_X F
-\frac{1}{\sqrt{2}}d\iota_X\iota_YC.
\eeq
This expression motivates the idea of a Courant bracket twisted by the 
3-gerbe $C$. Let a bracket with the suffix T be the twisted Courant bracket
defined by
\beq
[X+\xi,Y+\eta]_T = [X+\xi,Y+\eta]_C + \frac{1}{\sqrt{2}}\iota_Y\iota_XF
\eeq
This expression now agrees exactly  with the  algebra of diffeomorphisms
and gauge transformations found in the canonical theory.

One might
by concerned that the Jacobi identity fails for such expressions.
The Jacobiator of the Courant bracket is defined to be
\begin{multline}
J_C(X+\xi,Y+\eta,Z+\chi)=[[X+\xi,Y+\eta]_C,Z+\chi]_C
+[[Y+\eta,Z+\chi]_C,X+\xi]_C \\ + [[Z+\chi,X+\xi]_C.Y+\eta]_C.
\end{multline}
The Jacobiator vanishes if the Jacobi identity holds.
Explicit calculation shows that the Jacobiator of the Courant bracket is
\beq
J_C(X+\xi,Y+\eta,Z+\chi) = -d[\iota_X\iota_Zd\eta+\iota_Y\iota_Xd\chi+
\iota_Z\iota_Yd\xi] 
\eeq
Although the Jacobiator does not vanish, 
it is  an  exact two-form and thus has does not lead to a 
violation of the Jacobi identity when acting on the fields.

Similarly for the twisted Courant bracket. The Jacobiator
is then $J_T$ and is given by
\beq
J_T = J_C -\frac{1}{\sqrt{2}}\iota_Z\iota_Y\iota_X dF
\eeq
No physically meaningful violations of the Jacobi identity will 
be encountered as 
long as the field $F$ obeys its Bianchi identity, $dF=0$.
We therefore conclude that the correct generalization of the Lie bracket 
is the twisted Courant bracket and this reproduces the algebra of 
diffeomorphisms and gauge transformations. Perhaps the failure of the 
Jacobiator to vanish should be taken to mean that there is further structure 
here to be uncovered. It is tempting to speculate that this will be resolved
by understanding precisely what the sectioning condition really is.

Consider now the generalized tangent bundle. At each point in the 
four-dimensional space,  the vectors are in the ${\bf 4}$ of $SL(4)$
and the two-forms are  in the ${\bf 6}$ of $SL(4)$. Suppose 
that in an orthonormal frame the vectors have components $X^a$ 
and the two-forms have components $\xi_{ab}$. Furthermore,  suppose that the
tangent space is equipped with the norm 
$\bigl(\delta_{ab}, \frac{1}{2}(\delta^{ac}
\delta^{bd}-\delta^{ad}\delta^{bc})\bigr)$. 
Under the action of $SL(4)$, the vectors are mapped into vectors and 
the two-forms into two-forms in the obvious way and their norm will also be
preserved. Additionally, one can map vectors into two-forms using a three form, 
by the operation $\rho_C$. $C$ lies in a ${\bf 4}$ of $SL(4)$. 

There is an inverse operation which maps two-forms into vectors using 
a tri-vector, $B^{ijk}$. Under this map 
$(X^i,\xi_{jk})\rightarrow (X^i+B^{ijk}\xi_{jk},\xi_{jk})$.
$B^{ijk}$ is in another ${\bf 4}$ of $SL(4)$. 
The central observation is now that
these fields that are representations of $SL(4)$ can be unified into
a representation of $SL(5)$. By doing so, one has stopped treating the 
metric and three-form of M theory as different types of object, 
and put them onto a democratic footing, treating them both as different 
facets of a single geometric object. The generalized metric acts on a
${\bf 10}$ of $SL(5)$. $SL(5)$ can be decomposed into its $SL(4) \otimes U(1)$
subgroups. Under this decomposition the ${\bf 10}$ of $SL(5)$ is 
${\bf 6}_{-2} \otimes {\bf 4}_{3}$ being the vectors and two-forms 
respectively. The $U_1$ charges represent the scaling dimension of the
vectors and two-forms. 

The generalized metric is in the 
${\bf 24} \subset {\bf 10} \otimes \bar{\bf 10}$ of $SL(5)$. 
Under $SL(4)$, the ${\bf 24}$ decomposes as ${\bf 15}_0\oplus{\bf 4}
\oplus \bar{\bf 4} \oplus {\bf 1}$. the generalized metric  is in fact a 
parameterization of 
the symmetric space $SL(5)/SO(5)$, much as the reduced metric for general 
relativity was a parameterization of the symmetric space 
$SL_{\bar d}/SO({\bar d})$. The ${\bf 15}$ is the four-dimensional spatial metric,
and the ${\bf 4}$ is the three-form $C$. The $\bar{\bf 4}$ corresponds 
to the trivectors $B^{ijk}$, however the generalized metric is 
independent of these which 
reflects the local $SO(5)$ invariance of the system we are considering.

What have developed in this section is a description of the fields of our 
problem that show how it might be possible to construct a dynamical theory
using this unified description motivated by the structure of the U-duality 
group. In the next sections we carry out this program and discover that the 
theory has a global $SL(5)$ symmetry, which can be used to interchange 
the spatial metric and components of the three form. We also discover that
this picture is in no way dependent having compactified any of our spatial 
dimensions on a torus. It therefore is a tool which transcends our 
usual notions of U-duality.

What we have not done, and this will be the
subject of a future publication, is to exhibit the full power of the possibility
of mixing $x^i$ and $y_{ij}$ coordinates. Also, we fully expect that 
similar considerations will work in higher dimensions, although we are already 
aware of a number of complications to a straightforward extension our treatment.
This too will be the subject of a future publication.

\section{Canonical M-Theory Revisited}

Now that we have the appropriate constituent elements, we can attempt to
reconstruct the dynamical theory out of the generalized metric.
Following the canonical Hamiltonian formalism, there are two
types of the term. The kinetic terms which are essentially the momentum squared
and the potential terms given only in terms of the fields and their spatial
derivatives.

Let us begin with the potential terms. The potential in terms of the
generalized metric will be given by:

\begin{multline}
V = \gamma^{1/2} \Biggl(\frac{1}{12} M^{MN} (\partial_M M^{KL})( \partial_N
M_{KL} ) - {\frac{1}{2}} M^{MN} (\partial_N M^{KL}) (\partial_L M_{MK}) \\
+\frac{1}{12}   M^{MN} (M^{KL} \partial_M M_{KL})(M^{RS} \partial_N M_{RS}) 
+\frac{1}{4}    M^{MN} M^{PQ}(M^{RS} \partial_P M_{RS}) (\partial_M M_{NQ}) 
\Biggr)\,  
\end{multline}
where $\partial_M = \bigl(\frac{\partial}{\partial x^a},
\frac{\partial}{\partial y_{ab}}\bigr)$.
$V$ is then evaluated using the definition of the generalized
metric $M_{MN}$ described in section four by equations (\ref{genmet}) and
(\ref{genmet1}), in terms of the spatial metric and three form with 
the assumption 
that $\partial_y =0 $. We use just the spatial part of the metric in the 
computation of $V$ since we are only allowing spacelike duality transformations.
in the synchronous gauge, this means replacing $g_{ab}$ by $\gamma_{ij}$.
After a long and careful calculation, the result, up to a total derivative, is
\beq
V = \gamma^{1/2}(R(\gamma) - {\frac{1}{48}} F^2) \, .  
\eeq
This is exactly the potential term we had for eleven dimensional
supergravity. Note, that at first sight $V$  appears to stand 
little chance of being related to the Ricci 
scalar since it does not contain any terms second order in 
derivatives (as compared with 
terms quadratic in first derivatives). However the key is to take 
the usual Einstein Hilbert 
term and integrate by parts any terms second order in derivatives. 
This produces a Lagrangian
of the type given in $V$. This is also why the equivalence is only
up to surface terms. It should also be noted that $V$ is thus a $SL(5)$
scalar times $\gamma^{1/2}$. Strictly speaking this means that only $SL(5)$
transformations that preserve $\gamma$ are invariances of $V$. It is however 
probably the case, that once a proper section condition is found, one
will be able to extend the invariance to include those where $\gamma$ is 
varying. That is because the factor of  $\gamma^{1/2}$ more properly 
belongs with the $d^\dd x$
to create an invariant volume measure. We currently lack a formulation
that includes an invariant volume measure on both the $x$ and $y$
co-ordinates. 

It is also worth making the comment here that the first guess for such a
term would be the Ricci scalar of the generalized metric. This does
not give the right answer. Apart from numerical factors not working
out, crucially, one does not obtain the gauge invariant field strength
$F$ from $C$. This is a consequence of the traditional Ricci scalar
being constructed to be a scalar under diffeomorphisms. The reason is
as we have  seen in
the section three where we described the constraint algebra,
when there is a 
nontrivial C field, the constraints generate more than just the
diffeomorphisms. This also probably related to the fact that the beta function
of the doubled string theory \cite{berman2} does not give the usual
Ricci curvature of the doubled space.

Thus one suspects we need a generalized notion of curvature that
respects all of the symmetries. The recent work of Hohm, Hull and Zwiebach
\cite{Hohm:2010pp}\
 reaches similar conclusions for the doubled geometry of the
string. There they use insights from string field theory. Here we
simply have guessed the answer and checked it reduces to the usual
result once the dependence of the $y$ co-ordinates is removed.
Removing the $y$ dependence is essentially our choice of how to take a
section in the generalized space. The rules for how to pick a section more
generally are something that needs to be explored. Nevertheless it is
clear that this choice should produce the usual description of supergravity, 
and it does.

It is also important to note that it is equivalent only up to a total
derivative term. This is because to show the equivalence we had to
integrate the usual Einstein-Hilbert term by parts to remove the terms
that were second order in derivatives. Our potential term only
contains first derivatives on the fields so this trick of removing total
derivatives was necessary
to stand any chance of there being an equivalence between the two.

This suggests that there may be terms second order in derivatives that
can complete the above potential to really give something like a
generalized scalar curvature.

We are thus now left with the kinetic pieces. In terms of the {\it{generalized metric}}, $M_{IJ}$  
we find the following simple kinetic term,
\beq
T=- \gamma^{1/2} \frac{1}{12}(  ({\it tr}\ \dot M^{-1} \dot M) + ({\it tr}\ M^{-1}\dot M)^2 ) \label{eq:ke}
\eeq
and we discover when this is evaluated in terms of the original fields 
that this is identical to the kinetic term for $g_{ij}$ and $C_{ijk}$ 
in the usual canonical formalism given in section three. Of course  to see 
the equivalence  one has to take the kinetic term for M theory and 
replace the momenta by the time derivatives of the fields. In (\ref{eq:ke})\
the second term is the piece that is required to reproduce the 
``timelike'' direction for the trace of the metric. If the $V=0$
we see that considerations that parallel section two show that
the metric is just geodesic motion on the coset space $SL(5)/SO(5)$.

Thus we can reproduce the entire  Hamiltonian in a way has a manifest 
global  $SL(5)$ invariance
simply as $T+V$, apart from the remote possiblity of a  problem
 with factors of $\gamma$.

Lastly, the fact that the generalized geometry construction of $V$ reproduces
$R-\frac{1}{48}F^2$ indicates that it might be possible to include timelike
directions within this formulation. Under such circumstances, if the 
four dimensions are Lorentzian, then the coset space becomes
$SL(5)/SO(3,2)$, \cite{Hull:1998br}. However, since our treatment 
includes the conventional duality picture, the possibility of having complex 
fluxes and also strange signatures with multiple timelike directions 
for spacetime, \cite{Hull:1998ym}\ discourages us from pursuing 
such a possibility energetically
at present.

\section{Conclusions and Speculations}

First let us restate the main result. The Hamiltonian of gravity 
and a three form potential has been reformulated using generalized geometry 
in a manifestly $SL_5$ invariant way with no dimensional reduction of 
the theory.

We expect a similar treatment to be possible to make manifest 
the higher dimensional duality groups of M-theory. Of course it will 
be very interesting to see using this approach what happens for $E_8$ and
beyond.

It has been a very pleasant exercise to see the utility and naturalness 
of generalized geometry in this approach, without this geometric 
perspective such a 
construction would not have been possible. One can also imagine 
that the dynamics 
described here may be useful in providing some notion of 
curvature in generalized geometry.

In this regard one aspect immediately come to the fore. 
First, the equivalence was only 
shown up to boundary terms. In formulating a proper curvature 
these neglected terms may 
well be important. Physically these terms would also be interesting 
in discussing things 
like black hole thermodynamics where the usual boundary 
terms (ie Gibbons Hawking terms) 
provide a useful way of evaluating the free energy of a black hole. 
Whether the 
boundary terms can also be written in a manifestly duality 
invariant way using the 
generalized metric will be crucial to understanding the 
connection between black holes and 
duality groups in supergravity.

The boundary terms also allude to bigger issues regarding
 topological restrictions in 
this construction. Essentially we have ignored
 them and so the duality group should 
be viewed as a solution generating symmetry group for supergravity 
rather than as an exact equivalence of the theory in those backgrounds. 
A detailed understanding of winding modes and 
quantization of charges should restrict the 
duality group to being integer valued and then there 
should be equivalence just as for 
T-duality.

Another open question is to determine the section condition. Our choice here is 
certainly not the only one and in particular the topological 
restrictions on how to implement this choice will be crucial for the 
construction of U-folds.
Related to this is how one should construct the measure over the full
generalized space. That is we have worked with densities throughout
that need to be supplemented by $\sqrt{det(\gamma)}$ factors in
integrals. Since through our section condition we have removed the
need to integrate over the $y$ co-ordinates this is not an issue but a
full duality invariant treatment would require and duality invariant
measure over the full space.

As indicated in the body of the text, incorporating time in this scheme will 
be difficult since the dualities involving timelike directions 
are more exotic and yet 
it seems that this should be possible.

Other ways in which this approach might prove useful is to try and 
construct the higher order correction 
terms in curvature. Essentially the idea would be to use the duality 
group to restrict these terms. This has been useful in IIB string theory 
using the S-duality and it seems that a similar approach 
may be useful in M-theory.
Finally of course, it will be interesting to see how fermions fit into 
this picture. All of these topics we intend to return to in future works.

\section{Acknowledgements}

DSB wishes to thank DAMTP, University of Cambridge for continued 
hospitality and is supported
in part by the Queen Mary STFC rolling grant ST/G000565/1.
We would like to thank the Mitchell family for their generous hospitality
at Cook's Branch Nature conservancy where some of this work was carried out.
MJP would like to thank Mitchell foundation and Trinity College Cambridge 
for their support. We are grateful to the following poeple who made comments on 
aspects of the manuscript afer the initial preprint appeared: Hadi Godazgar, Chris Hull, 
 Bernard Julia, Emanuel Malek and Hermann Nicolai.

\end{document}